\documentclass[pra,twocolumn]{revtex4}

\usepackage{amssymb, amsmath, amsthm}
\usepackage{hyperref}
\usepackage{times}

\newcommand{\ket}[1]{| #1 \rangle}
\newcommand{\bra}[1]{\langle #1 |}

\newcommand{\oper}[2]{| #1 \rangle \langle #2 |}
\newcommand{\cwinprod}[2]{\langle \boldsymbol{#1} , \boldsymbol{#2} \rangle}
\newcommand{\cwinprodarg}[3]{\langle \boldsymbol{#1} , \boldsymbol{#2}(#3) \rangle}

\DeclareMathOperator{\Tr}{Tr}

\newtheorem{thm}{Theorem}
\newtheorem{lemma}{Lemma}

\newtheorem{prop}{Proposition}

\begin{document}

\title{Compatibility of subsystem states and convex geometry}
\author{William Hall}
\affiliation{Department of Mathematics, University of York, Heslington, York YO10 5DD, U.K.}
\email{wah500@york.ac.uk}

\begin{abstract}
The subsystem compatibility problem, which concerns the question of whether a set of subsystem states are compatible with a state of the entire system, has received much study. Here we attack the problem from a new angle, utilising the ideas of convexity that have been successfully employed against the separability problem. Analogously to an entanglement witness, we introduce the idea of a compatibility witness, and prove a number of properties about these objects. We show that the subsystem compatibility problem can be solved numerically and efficiently using semidefinite programming, and that the numerical results from this solution can be used to extract exact analytic results, an idea which we use to disprove a conjecture about the subsystem problem made by Butterley et al.\ [Found. Phys. \textbf{36} 83 (2006)]. Finally, we consider how the ideas can be used to tackle some important variants of the compatibility problem; in particular, the case of identical particles (known as $N$-representability in the case of fermions) is considered.
\end{abstract}
\maketitle

\section{Introduction}
The fact that quantum mechanical particles can be entangled with each other leads to a notion of \emph{compatibility} of quantum states. Given a finite system of quantum mechanical particles, the reduced states of a subset of these particles must arise from the partial trace of some density state of the entire system, and as a result the subsystem states cannot be arbitrary positive operators of trace one. A simple example comes from the `monogamy of entanglement': If two particles A and B are in a maximally entangled state, any other particle C cannot be entangled at all with either A or B \cite{CKW00, OsV06}. 

There are many variants of this problem, but the form which we will be most concerned with in this paper is the following  \cite{BSS}: \emph{Given states of all proper subsystems of a multipartite quantum system, what are the necessary and sufficient conditions for these subsystem states to be compatible with a single state of the entire system?}. A number of techniques and different approaches have been used to attack this problem, and many partial results have been obtained in recent years. However, much of the current works on the problem have concentrated on less general forms of the problem, and also on the \emph{spectral} properties of the involved states.

Another problem that has been the topic of even greater study in quantum mechanics is the \emph{separability problem}, of trying to decide when a multipartite quantum state is \emph{entangled} or \emph{separable} (not entangled). One approach to this problem is to use the fact that the set of separable states is a \emph{convex} set, and to use all of the geometric facts that follow from convexity. This turns out to be a very fruitful approach, and has led to the development of many powerful mathematical tools, most notably the idea of an \emph{entanglement witness}, for attacking the separability problem (see \cite{HHHBk} for a review).

What we aim to do here is to utilise the idea of convexity to attack the compatibility problem. We will show that in the correct mathematical framework the set of compatible subsystem states can be viewed as a convex set. We will define the notion of a \emph{compatibility witness}, which can be used to show when a set of subsystem states is \emph{not} compatible with a state for the entire system, and we use these to form the first necessary and sufficient condition for the compatibility problem. However, much as is the case with entanglement witnesses, without a full characterisation of compatibility witnesses, in practice this theorem is not of any operational use. Nonetheless, we do prove a number of interesting results about the nature of compatibility witnesses.

We will show that the ideas of \emph{semidefinite programming} \cite{BL96, BLBk} can be used to numerically solve the compatibility problem. The inspiration for this lies behind similar applications of semidefinite programming to the separability problem \cite{DPS02,DPS,EHOC04,BrV04a,BrV04b}. We will give a brief overview of semidefinite programming and illustrate how it can be applied to the compatibility problem. Furthermore, we will show how analytic results can be obtained from the numerical results of our solution, and we use it to disprove a conjecture of Butterley et al.\ \cite{BSS} concerning the necessary and sufficient conditions for the compatibility problem involving three qubits (the simplest non-trivial case).

There are other variants of the problem that we will consider after the main problem. An important situation is one where we only have partial knowledge of the subsystem states. Another is the case where we have $n$ identical particles (bosons or fermions); this problem is of particular importance in quantum chemistry in determining the ground state energies of a system of identical particles \cite{Maz06}. We will be able to give solutions to both of these variants using a modification of the semidefinite program used to solve the main problem. 

\section{The compatibility problem and previous results}

\subsection{Statement of the problem}

Let us first establish some nomenclature. For any Hilbert space $\mathcal{K}$, let $\mathcal{B}(\mathcal{K})$ be the set of bounded operators on $\mathcal{K}$, and let $\mathcal{D}(\mathcal{K})$ be the space of density states (i.e. positive operators of trace 1). 

Let $\mathcal{H}_i$ ($i=1,\ldots,n$) be Hilbert spaces with finite dimension $d_i$. Define $\mathcal{H} = \mathcal{H}_1 \otimes \mathcal{H}_2 \otimes \ldots \otimes \mathcal{H}_n$ to be the Hilbert space for an $n$-partite quantum system, and define $N = \{ 1, \ldots, n \}$. A set of states on every proper subsystem then corresponds to a set of operators $\rho_A \in \mathcal{B}(\otimes_{i \in A} \mathcal{H}_i)$ for every $A \subset N$.

The compatibility problem asks the following question: given $\rho_A$ for all $A \subset N$, what are the necessary and sufficient conditions for these states to compatible with a state $\rho \in \mathcal{H}$, i.e. 
\begin{equation} \rho_A = \Tr_{N \setminus A} \label{com} \rho \end{equation}
where $\Tr_X$ represents the partial trace on systems $\mathcal{H}_i$ for all $i \in X$. We say the set of subsystem states are \emph{incompatible} if no such state can be found.

Some obvious compatibility conditions arise from the fact that two states describing overlapping systems must yield the same reduced states when the overlap is traced out. Formally, we can state this as follows: if $A,B,C \subset N, A \cap B = A \cap C = \emptyset$, then
\begin{equation} \Tr_B(\rho_{A \cup B}) = \Tr_C(\rho_{A \cup C}). \label{trace} \end{equation}

Even in the simplest case of three qubits, there is no known solution to this problem, and there is no analytic way of determining whether an arbitrary set of subsystem states is compatible with a full description of the system (or not).

\subsection{Summary of current results} \label{ss_prev_res}

There are very few known necessary criteria for the general form of the compatibility problem as stated above. The following theorem is an example of such a condition:

\begin{thm} \label{com_conj} Let $\rho_A = \Tr_{N \setminus A} \rho$ (where $A \subset N=\{1,\ldots,n\}$) be the reduced states of a $n$-party state $\rho$. Then, defining  
\begin{equation} \Delta = \sum_{A \subset N} (-1)^{|A|} \rho_A \end{equation}
(where $\rho_A$ here is padded out with identities on the traced out systems), then if $n$ is odd, $\Delta \geq 0$. If all of the systems are qubits, then $\Delta \leq 1$ also.
\end{thm} 

This was proved in the case of three qubits in \cite{BSS}. and the result in its full generality can be inferred from \cite{Hall}. Furthermore, through a study of the classical analog of the compatibility problem, it was conjectured in \cite{BSS} that in the case of three qubits, the condition $0 \leq \Delta \leq 1$ is not only necessary for subsystem state compatibility, but also sufficient. The three party case is also studied in detail in \cite{HZG05}, where a number of necessary conditions on the spectra of any three party state and its two-party reduced states are produced.

The fact that we only have necessary conditions reflects the difficulty in the problem. A set of subsystem states can be proved to be incompatible if they violate any of these necessary conditions. However, the only sufficient conditions we have amount to checking whether the subsystem states are compatible with a specific state $\rho \in \mathcal{B}(\mathcal{H})$, and checking these conditions for all possible states is not operationally feasible.

There do however exist many results in more constrained versions of the problem. One question that has been particularly studied is the following: \emph{What are the conditions on the spectra of the one-particle reduced states for the full state to have a given spectrum?} For example, a pure state has a particular spectrum, and the necessary and sufficient conditions on the spectra of the one-particle states for them to be the reduced states of a pure state are derived for any number of qubits by Higuchi et al. in \cite{HSS03}, for three qutrits by Higuchi in \cite{Hig03}, and for two qubits and a 4-dimensional system by Bravyi in \cite{Brav04}. A set of necessary conditions were proved for any number of arbitrary systems by Han et al. in \cite{HZG04}. Finally, the most general form of the problem was solved by Klyachko \cite{Kly04}, using methods of symplectic geometry similar to those used by Klyachko to solve a long-standing conjecture by Horn, who conjectured a set of inequalities answering the question `What are the possible spectra of a sum of Hermitian matrices with given spectra?'. Related to this is the work of Daftuar and Hayden \cite{DH05}, in which the possible spectra of a one-party reduced state of a bipartite state are determined.

As mentioned already, most of this work concentrates on establishing relationships between the spectra of the reduced and (where one can exist) the total states. In this work we intend to move away from this focus and concentrate more on the geometric aspects of the problem at hand.

Other works that establish analytic results about the problem include the work of Christandl and Mitichison \cite{CM05}, in which a connection between this problem and the representation theory of the symmetric group is obtained, and the work of Jones and Linden \cite{JL05}, in which a system of qubits is considered, and it is shown that if we are given all the reduced states of more than half the number of qubits, there is almost always at most one pure state compatible with these subsystem states.

Another important variant of the problem is the case of identical particles. This extra condition places more constraints on the density state for the entire system, as pure states of bosons (fermions) are symmetric (antisymmetric) under exchange of any two particles. A compatibility problem that has received particular attention is one that has become known as the $N$-representability problem - this is the case when we have $N$ identical fermions. The necessary conditions on the one-party reduced states was solved by Coleman:

\begin{thm}[Coleman \cite{Col63}, Corollary 4.3A] A density matrix $\rho$ is the reduced one-party state of a system of $N$ fermions if and only all of its eigenvalues are bounded above by $1/n$. \end{thm}

The necessary and sufficient conditions on the two-party reduced state however are still not known, and have been of great interest in the field of quantum chemistry, since knowing the conditions on the two-party states would allow the calculation of the ground state energy of an $N$-fermion Hamiltonian involving 2-body interactions only from the 2-party state and an effective 2-body Hamiltonian. This is still a topic of active research (see \cite{Maz06} and references therein).

Finally, there has been some research on the complexity of the compatibility problem. Specifically, it has been shown that when we have partial knowledge of the reduced states (specifically, a number of reduced states that is polynomial in the number of quantum systems, each describing a number of systems less than some constant), the corresponding compatibility problem is QMA-complete \footnote{The complexity class QMA is the quantum computing analog of NP. Roughly speaking, a problem is in QMA if a solution to a problem can be encoded into a quantum state of a number of qubits polynomial in the problem size, and verified by a quantum circuit (verifier) in polynomial time. See \cite{AhN02} for a more precise definition.} and hence also NP-hard. It was also shown very recently in \cite{LCV06} that the specific $N$-representability problem where we are given the 2-party reduced state is also in the complexity class QMA-complete. In this paper we will not be concerned with detailed analysis with respect to complexity classes, but we will discuss the computational efficiency and complexity of the algorithms involved.

\section{Convexity and the compatibility problem}

In this section we will use the ideas of convexity on the compatibility problem, and introduce the notion of a compatibility witness, a tool that will allow us to detect sets of subsystem states that are not compatible with an overall state.

\subsection{Compatible states as a convex set}

Let us define $\mathcal{H}^{(k)} = \otimes_{j \in N, j \neq k} \mathcal{H}_j$ (i.e. system $k$ removed from the tensor product), and let us define $\mathcal{O} \subset \oplus_{i=1}^n \mathcal{B}(\mathcal{H}^{(i)})$ by
\begin{equation}
\mathcal{O} = \{ \ \boldsymbol{O}=(O_1, \ldots, O_n) \ | \ O_i \in \mathcal{B}(\mathcal{H}^{(i)}), \ O_i^\dagger = O_i \}.
\end{equation}
We define an inner product on this space by
\begin{equation} \cwinprod{A}{B} = \sum_{i=1}^n \Tr(A_iB_i) \label{cw_ip}. \end{equation}
This space is a natural space to define the set $C \subset \mathcal{O}$ by 
\begin{equation}
C = \{ \boldsymbol{\rho} = (\rho_1,\ldots,\rho_n) \ | \ \rho_j = \Tr_j \rho \textrm{ for some } \rho \in \mathcal{D}(\mathcal{H}) \}. 
\end{equation}
This is the set of compatible subsystem states, or more precisely, the set of vectors of the $n$ states of $n-1$ systems obtained from the partial trace of a full state $\rho$ ($\rho_i$ corresponds to the state on $\mathcal{H}^{(i)}$). The states of a smaller number of systems can then be obtained from partially tracing an appropriate $(n-1)$-system state. 

The compatibility problem can now be cast in the following form: Let $\boldsymbol{\sigma} = (\sigma_1,\ldots,\sigma_n)$ be a vector of the $n$ states of $n-1$ systems. We are then trying to ascertain whether $\boldsymbol{\sigma}$ lies in the set $C$. This is rather similar to the separability problem - given a state $\rho$, we are trying to ascertain whether it belongs to the set of separable states. This set is convex and compact, and we will prove the same fact about $C$.

\begin{prop} C is a convex, compact subset of $\mathcal{O}$. \end{prop}

\begin{proof} We will first prove $C$ is convex. Let $\boldsymbol{\rho},\boldsymbol{\sigma} \in C$. Then $\rho_i = \Tr_i \rho, \sigma_i = \Tr_i \sigma$ for some quantum states $\rho,\sigma \in \mathcal{B}(\mathcal{H})$. Let $\alpha \in [0,1]$. Then
\begin{equation} \alpha \rho_i + (1-\alpha) \sigma_i = \Tr_i(\alpha \rho + (1-\alpha) \sigma) \end{equation}
and since $\alpha \rho + (1-\alpha) \sigma \in \mathcal{D}(\mathcal{H})$, $\alpha\boldsymbol{\rho} + (1-\alpha)\boldsymbol{\sigma} \in C$.

Now we will prove $C$ is compact. Define the function $R: \mathcal{B}(\mathcal{H}) \to \oplus_{i=1}^n \mathcal{B}(\mathcal{H}^{(i)})$ by
\begin{equation} R(\rho) = (\Tr_1 \rho, \ldots, \Tr_n \rho). \end{equation}
By definition, $R(\mathcal{D}(\mathcal{H})) = C$. Since the trace is a continuous function, $R$ is a continuous function. Furthermore, $\mathcal{D}(\mathcal{H})$ is a closed and bounded set, and so by the Heine-Borel theorem \cite{SutBk}, it is compact. Finally, the continuous image of a compact subset is itself compact, and hence $R(\mathcal{D}(\mathcal{H})) = C$ is compact. \end{proof}

\subsection{Compatibility witnesses}

The idea of a entanglement witness arises from the Hahn-Banach theorem. An important geometric fact that follows from this theorem is the following \cite{HHH96, EdBk}:

\begin{thm} Let $C_1,C_2$ be disjoint convex closed sets in a real Banach space, and let $C_1$ be compact. Then there exists a continuous functional $f$, and $a \in \mathbb{R}$ such that, for every $c_1 \in C_1, c_2 \in C_2$, $f(c_1)<a\leq f(c_2)$. \end{thm}

We can use this theorem to prove a result about elements of $\mathcal{O}$ not in $C$ analogous to the result about entangled states in \cite{HHH96}:

\begin{prop} Let $\boldsymbol{\sigma} \in \mathcal{O}$, but $\boldsymbol{\sigma} \notin C$. Then there exists $\boldsymbol{W} = (W_1, \ldots, W_n) \in \mathcal{O}$ such that $\cwinprod{W}{\sigma} <0$, and $\cwinprod{W}{\rho} \geq 0$ for all $\boldsymbol{\rho} = (\rho_1,\ldots,\rho_n) \in C$. \end{prop}

\begin{proof} The one-element set $S(\boldsymbol{\sigma}) = \{ \boldsymbol{\sigma} \}$ and $C$ are convex compact (and hence closed) subspaces of the real Banach space $\mathcal{O}$. Since also $S(\boldsymbol{\sigma}) \cap C = \emptyset$,  we can apply the above theorem. Hence there exists a linear functional $g: \mathcal{O} \to \mathbb{R}$ and $a \in \mathbb{R}$ such that 
\begin{equation} g(\boldsymbol{\sigma}) < a \leq g(\boldsymbol{\rho}) \end{equation}
for all $\boldsymbol{\rho} \in C$. By the Riesz-Fr\'{e}chet representation theorem, 
\begin{equation} g(\boldsymbol{X}) = \sum_{i=1}^n \Tr(V_i X_i) \end{equation}
where $\boldsymbol{V}=(V_1, \ldots, V_n) \in \mathcal{O}$. Now, defining $\boldsymbol{W} = \boldsymbol{V} - a\boldsymbol{I}/n$, where $\boldsymbol{I}=(\boldsymbol{1}_1/d_1, \ldots, \boldsymbol{1}_n/d_n)$, we obtain that
\begin{equation}
\sum_{i=1}^n \Tr(W_i \sigma_i) <0 \leq \sum_{i=1}^n \Tr(W_i \rho_i)
\end{equation}
which is the required result. \end{proof}

This result immediately leads to the following theorem:

\begin{thm} \label{cw_iff} Let $\boldsymbol{\sigma} \in \mathcal{O}$. Then the $(n-1)$-party states represented by $\boldsymbol{\sigma}$ are compatible with a state of the full system (i.e. $\boldsymbol{\sigma} \in C$) if and only if $\cwinprod{W}{\sigma} \geq 0$ for all $\boldsymbol{W} \in \mathcal{O}$ such that $\cwinprod{W}{\rho} \geq 0$ for all $\boldsymbol{\rho} \in C$. \end{thm}

We call any $\boldsymbol{W}$ satisfying $\cwinprod{W}{\rho} \geq 0$ for all $\boldsymbol{\rho} \in C$ a \emph{compatibility witness}, and we will denote the set of compatibility witnesses by $\mathcal{W}$. If $\cwinprod{W}{\rho} <0$, the subsystem states $\boldsymbol{\rho}$ are incompatible, and we say $\boldsymbol{W}$ \emph{detects} the incompatibility in $\boldsymbol{\rho}$.

This is the first example of a necessary and sufficient condition for the compatibility problem, but (rather like the similar theorem for entanglement witnesses) it is not immediately operationally useful, because it requires a complete characterisation of the entire set of compatibility witnesses. 

\subsection{Properties of compatibility witnesses}

To make the introduction to compatibility witnesses easier, we will prove a few results about some properties of compatibility witnesses. The first shows us how to jump between a compatibility witness and a positive operator characterising this witness. 
In all that follows this point, if $A \in \mathcal{B}(\mathcal{H}^{(i)})$, and $B \in \mathcal{B}(\mathcal{H}_i)$, then the operator $A \otimes B$ should be interpreted as being a member of $\mathcal{B}(\mathcal{H})$ rather than $\mathcal{B}(\mathcal{H}^{(i)}) \otimes \mathcal{B}(\mathcal{H}_i)$ (for example, if $n=3$, and $A = A_1 \otimes A_3 \in \mathcal{B}(\mathcal{H}^{(2)})$, then $A \otimes B$ should be interpreted as $A_1 \otimes B \otimes A_3 \in \mathcal{B}(\mathcal{H})$).

\begin{prop} \label{cw_pos} $\boldsymbol{W}$ is a compatibility witness if and only if the operator $p(\boldsymbol{W}) \in \mathcal{B}(\mathcal{H})$ defined by
\begin{equation} p(\boldsymbol{W}) = \sum_{i=1}^n W_i \otimes \boldsymbol{1}_i, \label{cw_full} \end{equation}
where $\boldsymbol{1}_i$ is the identity operator on $\mathcal{H}_i$, is positive. \end{prop}

\begin{proof} Let $\boldsymbol{\rho} \in C$ be compatible with a state $\rho$. We note that
\begin{equation}
\sum_{i=1}^n \Tr(W_i \rho_i) = \Tr \left[ \sum_{i=1}^n ( W_i \otimes \boldsymbol{1}_i) \rho \right] = \Tr (p(\boldsymbol{W})\rho)
\end{equation}
and so the result follows (i.e. if one side is greater than zero, so is the other). \end{proof}

However, constructing non-trivial entanglement witnesses is not as simple as taking $W_i$ to be positive for all $i=1,\ldots,n$, as then $\Tr(W_i \sigma_i) \geq 0$ for any positive $\sigma_i$ and so the compatibility witness $\boldsymbol{W}$ never detects any incompatibility of subsystem states.

We should note here that the function $p$ is not a one-to-one function: there exist $\boldsymbol{O} \in \mathcal{O}$ (not necessarily a compatibility witness) such that $p(\boldsymbol{O}) = 0$, and hence $\boldsymbol{W^\prime} = \boldsymbol{W} + \boldsymbol{O}$ is a compatibility witness such that $p(\boldsymbol{W^\prime}) = p(\boldsymbol{W})$. We will return to this point later.

The structure of $p(\boldsymbol{W})$ also implies the following result:

\begin{prop} \label{prop_t} Let $\boldsymbol{W} \in \mathcal{O}$ (not necessarily a compatibility witness), and let $T_i \in \mathcal{B}(\mathcal{H}_i)$, with $\Tr(T_i) = 0$. Then, defining $T=T_1 \otimes \ldots \otimes T_n$, $\Tr(p(\boldsymbol{W})T)=0$. \end{prop}

\begin{proof} Define $T^{(i)} = \otimes_{k\neq i} T_k$. We see that
\begin{eqnarray}
\Tr(p(\boldsymbol{W})T) &=& \Tr \left(\sum_{i=1}^n (W_i \otimes \boldsymbol{1}_i)T \right) \\
&=& \sum_{i=1}^n \Tr(W_iT^{(i)})\Tr(T_i) = 0, 
\end{eqnarray}
establishing the result. \end{proof}

This can be reduced to a finite number of conditions by taking a basis for the subspace of traceless matrices in each Hilbert space $\mathcal{H}_i$, and then forming the tensor product of these bases (this yields $\prod_{i=1}^n (d_i^2-1)$ conditions). 

We can also consider this problem in reverse: given a positive operator $Z \in \mathcal{B}(\mathcal{H})$, we may want to ascertain whether this operator is of the form $Z=p(\boldsymbol{W})$. This is all encapsulated in the following proposition:

\begin{prop} \label{prop_t2} Let $Z \in \mathcal{B}(\mathcal{H})$ be positive, such that $Tr\left[Z(T_1 \otimes \ldots \otimes T_n)\right]=0$ if $\Tr(T_i)=0$. Then $Z=p(\boldsymbol{W})$ for some compatibility witness $\boldsymbol{W}$.  \end{prop}

\begin{proof} Let $B_{k,m}$ ($m=1,\ldots,d_k^2$) be a basis of Hermitian operators for $\mathcal{B}(\mathcal{H}_k)$ satisfying
\begin{equation} \label{basis_o} B_{k,m} = \frac{\boldsymbol{1}_k}{d_k}; \ \Tr(B_{k,m}) = \delta_{1 m}; \ \Tr(B_{k,m} B_{k,n} ) = \frac{1}{d_k} \delta_{mn} \end{equation}
and define
\begin{equation} B_\mathbf{m} = \otimes_{k=1}^n B_{k,m_k}. \label{basis} \end{equation}
Define $D = \dim \mathcal{H} = \prod_i d_i$. The set $\{ B_\mathbf{m} \ | \ m_i=1,\ldots,d_i^2; \ i=1,\ldots,n \}$ forms a basis for $\mathcal{B}(\mathcal{H})$, and so we can write
\begin{equation} Z = \sum_{\mathbf{m} \in I} z_\mathbf{m} B_\mathbf{m}. \end{equation}
where $I = \{ \mathbf{m} \ | \ m_i=1,\ldots,d_i^2; \ i=1,\ldots,n \}$, and (from the orthogonality conditions (\ref{basis_o})), $z_\mathbf{m} = D \Tr(ZB_\mathbf{m})$. Now define the set
\begin{eqnarray} 
I_C &=& \left\{ \mathbf{m} \ | \ m_i=1,\ldots,d_i^2, \ i=1,\ldots,n; \right. \nonumber \\ && \left. m_k=1 \textrm{ for some } k=1,\ldots,n \right\} . \label{is} 
\end{eqnarray}
For $Z$ satisfying the assumptions in the statement of the proposition, we have that $z_\mathbf{m} = \Tr(ZB_\mathbf{m})=0$ for any $\mathbf{m} \notin I_C$, as these basis elements are a tensor product of traceless operators. Hence $Z$ can be written in the form 
\begin{equation}
Z = \sum_{\mathbf{m} \in I_C} z_\mathbf{m} B_\mathbf{m}. \label{cw_basis}
\end{equation}
When $\mathbf{m} \in I_C$, the basis element $B_\mathbf{m}$ is of the form $B_i \otimes \boldsymbol{1}_i$ for some $i=1,\ldots,n$, and so $Z$ is of the form $\sum_i W_i \otimes \boldsymbol{1}_i$. Hence $Z=p(\boldsymbol{W})$ for some $\boldsymbol{W} \in \mathcal{O}$, and since $Z$ is positive, by Proposition \ref{cw_pos}, $\boldsymbol{W}$ is a compatibility witness. \end{proof}

Included in the above proof is a method of extracting a suitable $\boldsymbol{W} \in \mathcal{W}$ such that $Z=p(\boldsymbol{W})$. The operators $z_\mathbf{m} B_\mathbf{m}$ for $\mathbf{m} \in I_C$ in (\ref{cw_basis}) can be used to construct the operators $W_i$, where $i$ is determined by the component of $\mathbf{m}$ that equals 1. Where more than one component equals 1, the operator can contribute to any of the appropriate $W_i$. This freedom represents the fact that there is not a unique $\boldsymbol{W} \in \mathcal{W}$ satisfying $Z=p(\boldsymbol{W})$, but there are in fact many such witnesses.

Another method of finding an appropriate $\boldsymbol{W} \in \mathcal{W}$ from $Z$ is through the following proposition:

\begin{prop} \label{prop_cwtest} Let $Z = p(\boldsymbol{W})$ for some $\boldsymbol{W} \in \mathcal{W}$. Then 
\begin{equation} Z = -\sum_{A \subseteq N, A \neq \emptyset} \frac{(-1)^{|A|}}{d_A} Z_{N \setminus A} \otimes \boldsymbol{1}_A \label{cw_test} \end{equation}
where $d_A = \prod_{i \in A} d_i$, and $1_A = \otimes_{i \in A} 1_i$. \end{prop}

\begin{proof} Using the principle of inclusion/exclusion, we can express (\ref{cw_basis}) as follows:
\begin{eqnarray}
Z &=& \sum_{k=1}^n (-1)^{k+1} \sum_{A \subseteq N, |A|=k} \left( \sum_{ \substack{ \mathbf{m} \in I \\ m_l=1 \textrm{ when } l \in A} } z_\mathbf{m} B_\mathbf{m} \right) \\
&=& -\sum_{A \subseteq N, A \neq \emptyset} \frac{(-1)^{|A|}}{d_A} \Tr_A(Z) \otimes \boldsymbol{1}_A \\
&=& -\sum_{A \subseteq N, A \neq \emptyset} \frac{(-1)^{|A|}}{d_A} Z_{N \setminus A} \otimes \boldsymbol{1}_A
\end{eqnarray}
which is the required result. \end{proof}
Equation (\ref{cw_test}) allows us to express $Z$ in the form $\sum_{i=1}^n Z_i \otimes 1_i$, and so find a suitable $\boldsymbol{W}$. For example, in the case $n=3$, given $Z=p(\boldsymbol{W})$, a suitable $\boldsymbol{W}$ is given by
\begin{eqnarray}
W_1 &=& \frac{\boldsymbol{1}}{3d_1d_2d_3} - \frac{\boldsymbol{1}_1 \otimes Z_2 \otimes \boldsymbol{1}_3}{d_1d_3} + \frac{\boldsymbol{1}_1 \otimes Z_{23}}{d_1}; \\
W_2 &=& \frac{\boldsymbol{1}}{3d_1d_2d_3} - \frac{\boldsymbol{1}_1 \otimes \boldsymbol{1}_2 \otimes Z_3}{d_1d_2} + \frac{\boldsymbol{1}_2 \otimes Z_{13}}{d_2}; \\
W_3 &=& \frac{\boldsymbol{1}}{3d_1d_2d_3} - \frac{Z_1 \otimes \boldsymbol{1}_2 \otimes \boldsymbol{1}_3}{d_2d_3} +  \frac{Z_{12} \otimes \boldsymbol{1}_3}{d_3}.
\end{eqnarray}
We again see the freedom of choice in $\mathcal{W}$ here, as some of the terms could appear in more than one of the operators $W_i$, but we have chosen this symmetric expression. This method for establishing $\boldsymbol{W}$ from $p(\boldsymbol{W})$ will be important when we come to talk about using semidefinite programming on the compatibility problem.

\subsection{Optimal compatibility witnesses} \label{ss_ocw}

Given two compatibility witnesses $\boldsymbol{W}, \boldsymbol{W^\prime}$, if $\boldsymbol{W^\prime}$ detects all the incompatible subsystem states that $\boldsymbol{W}$ does, we can discard $\boldsymbol{W}$ and use $\boldsymbol{W^\prime}$ instead. More generally, we need a notion of an \emph{optimal} compatibility witness, a witness which cannot be replaced by another witness that detects as least as many incompatible subsystem states.

In \cite{LKCH00}, the notion of an optimal \emph{entanglement} witness is introduced. As a result, a number of the results and proofs here are similar to those given in \cite{LKCH00}, but with some differences due to the different structure of this problem.

Let us take compatibility witnesses $\boldsymbol{W}$ satisfying the normalisation constraint $\Tr(p(\boldsymbol{W}))=1$. We start with some definitions:
\begin{itemize}
\item A compatibility witness $\boldsymbol{W}$ is \emph{tangential} to $C$ if there exists $\boldsymbol{\rho} \in C$ such that $\cwinprod{W}{\rho} = 0$.
\item $\boldsymbol{W^\prime}$ is \emph{finer} than $\boldsymbol{W}$ if $\cwinprod{W}{\rho} < 0 \Rightarrow \cwinprod{W^\prime}{\rho} < 0$ (i.e. $\boldsymbol{W^\prime}$ detects at least as many incompatible subsystem states as $\boldsymbol{W}$).
\item $\boldsymbol{W}$ is an \emph{optimal} compatibility witness if, apart from $\boldsymbol{W}$ itself, no compatibility witness is finer than $\boldsymbol{W}$.
\end{itemize}

The normalisation constraint means that we are fairly comparing witnesses (under the above definition $\alpha\boldsymbol{W}$ is finer than $\boldsymbol{W}$ when $\alpha \neq 0$, even though they clealy detect the same set of incompatible subsystem states; in fact, without this normalisation, optimal compatibility witnesses would not exist).
 
The first observation of note here is that an optimal witness is also tangential: if
\begin{equation} \min_{\boldsymbol{\rho} \in C} \cwinprod{W}{\rho} = \lambda_{\min} > 0 \end{equation}
then the witness $\boldsymbol{W^\prime}$ defined by $W_i^\prime = W_i - \alpha \boldsymbol{1}_i / n$ is a finer compatibility witness, and is also tangential to $C$ at the point $\rho \in C$ that minimises the above expression. However, since
\begin{equation} \cwinprod{W}{\rho} = \Tr(p(\boldsymbol{W})\rho) \end{equation}
(where $\rho$ is a state compatible with $\boldsymbol{\rho}$), then $\lambda_{\min}$ is simply the minimal eigenvalue of $p(\boldsymbol{W})$. Hence we have the following:

\begin{lemma} $\boldsymbol{W}$ is tangential to $C$ if and only if $p(\boldsymbol{W})$ has a zero eigenvalue, and $\boldsymbol{W}$ is optimal only if $p(\boldsymbol{W})$ has a zero eigenvalue. \end{lemma}

Given a non-optimal compatibility witness $\boldsymbol{W}$, we may ask how to produce a finer witness $\boldsymbol{W^\prime}$ from it. Suppose that the minimum eigenvalue of $p(\boldsymbol{W})$ is $\lambda^W_{\min} > 0$, and let $\boldsymbol{P}$ be a compatibility witness such that $\cwinprod{P}{\rho} \geq 0$ for all $\rho \in \mathcal{O}$ (not just the compatible subsystem states), with maximum eigenvalue $\lambda^P_{\max}$. Then, by Proposition \ref{prop_t2}, the operator
\begin{equation} Z^\prime = p(\boldsymbol{W}) - \frac{\lambda^W_{\min}}{\lambda^P_{\max}} p(\boldsymbol{P}) \end{equation}
defines a compatibility witness $\boldsymbol{W^\prime}$ which is by definition finer than $\boldsymbol{W}$, since $\cwinprod{W^\prime}{\rho} = \cwinprod{W}{\rho} - \frac{\lambda^W_{\min}}{\lambda^P_{\max}}\cwinprod{P}{\rho} < \cwinprod{W}{\rho}$.

The rest of this section is dedicated to proving the converse of this, i.e.:

\begin{thm} \label{cw_opt_thm} A compatibility witness $\boldsymbol{W^\prime}$ is finer than $\boldsymbol{W}$ if and only if there exists $\boldsymbol{P}$ such that $\cwinprod{P}{\rho} \geq 0$ for all $\rho \in \mathcal{O}$ and 
\begin{equation} \boldsymbol{W} = (1-\epsilon)\boldsymbol{W^\prime} + \epsilon \boldsymbol{P} \label{cw_opt} \end{equation}
where $\epsilon \in (0,1)$. \end{thm}

Before we are in a position to prove this theorem, we need the following lemma, which we prove here first.

\begin{lemma} \label{cw_id} Let $\boldsymbol{\rho}\in \mathcal{O}, \notin C$. Let $\boldsymbol{I} = 1/D (d_1 \boldsymbol{1}_1, \ldots, d_n \boldsymbol{1}_n )$, the reduced states of the maximally mixed state on $\mathcal{H}$. Then there exists $x \in (0,1)$ such that $\boldsymbol{\rho}(x) = (1-x)\boldsymbol{I} + x\boldsymbol{\rho}$ is compatible with a quantum state $\rho(x) \in \mathcal{B}(\mathcal{H})$. \end{lemma}

\begin{proof} For all compatibility witnesses satisfying $\Tr(p(\boldsymbol{W}))=1$, we have $\cwinprod{W}{I}=1$. Define
\begin{equation} c = \min_{\boldsymbol{W}} \cwinprod{W}{\rho} \end{equation}
where the minimisation is over all compatibility witnesses $\boldsymbol{W}$ satisfying $\Tr(p(\boldsymbol{W}))=1$. Since $\boldsymbol{\rho} \notin C$, $c <0$. Also, $\min_{\boldsymbol{W} \in \mathcal{W}, \Tr(p(\boldsymbol{W}))=1} \cwinprodarg{W}{\rho}{x} = (1-x) + xc$, and because of Proposition \ref{cw_iff}, for $x\leq 1/(1-c)$, $\boldsymbol{\rho}(x)$ is compatible with an overall state. \end{proof}

The following lemma is very similar to Lemma 1 in \cite{LKCH00}, and hence the proof is almost identical, with some small changes. We state the proof here however for completeness.

\begin{lemma} \label{cw_lam} For a compatibility witness $\boldsymbol{W}$, let $D_{\boldsymbol{W}} = \{ \boldsymbol{\rho} \in \mathcal{O} \ | \ \cwinprod{W}{\rho} < 0 \}$, and for two witnesses $\boldsymbol{W}, \boldsymbol{W^\prime}$, define
\begin{equation} \lambda = \min_{\boldsymbol{\sigma} \in D_{\boldsymbol{W}}} \left| \frac{ \cwinprod{W^\prime}{\sigma} }{ \cwinprod{W}{\sigma} } \right| . \end{equation}
Suppose $\boldsymbol{W^\prime}$ is finer than $\boldsymbol{W}$, Then the following statements are true: 

(i) $\cwinprod{W}{\rho} =0 \Rightarrow \cwinprod{W^\prime}{\rho} \leq 0$. 

(ii) $\cwinprod{W}{\rho} < 0 \Rightarrow \cwinprod{W}{\rho} \geq \cwinprod{W^\prime}{\rho}$.  

(iii) $\cwinprod{W}{\rho} > 0 \Rightarrow \lambda \cwinprod{W}{\rho} \geq \cwinprod{W^\prime}{\rho}.$ 

(iv) $\lambda \geq 1$, with equality if and only if $\boldsymbol{W}=\boldsymbol{W^\prime}$. \end{lemma}

\begin{proof} (i) Suppose $\cwinprod{W}{\rho}=0$, and assume $\cwinprod{W^\prime}{\rho}>0$. Take $\boldsymbol{\sigma} \in D_{\boldsymbol{W}}$, and define $\boldsymbol{\rho}(x) = x\boldsymbol{\sigma} + (1-x)\boldsymbol{\rho}$, so that $\boldsymbol{\rho}(x) \in D_{\boldsymbol{W}}$ for $x \in (0,1)$. For small enough $x \in (0,1)$, we have $\cwinprodarg{W^\prime}{\rho}{x}>0$, contradicting $\boldsymbol{W^\prime}$ being finer than $\boldsymbol{W}$.

(ii) Suppose $\cwinprod{W}{\rho}<0$. Define $\boldsymbol{\sigma}=\boldsymbol{\rho} + |\cwinprod{W}{\rho}|\boldsymbol{I}$, so that $\cwinprod{W}{\sigma}=0$. (i) then implies the result.

(iii) Suppose $\cwinprod{W}{\rho} > 0$, and take $\boldsymbol{\sigma} \in D_{\boldsymbol{W}}$. Define $\boldsymbol{\mu} = \cwinprod{W}{\rho} \boldsymbol{\sigma} + |\cwinprod{W}{\sigma}|\boldsymbol{\rho}$, so that $\cwinprod{W}{\mu}=0$. Then (i) implies that $\cwinprod{W^\prime}{\mu} < 0$ i.e. $|\cwinprod{W}{\sigma}|\cwinprod{W^\prime}{\rho} \leq |\cwinprod{W}{\rho}| \cwinprod{W^\prime}{\sigma}$. Dividing both sides by $|\cwinprod{W}{\sigma} | >0$ and $\cwinprod{W}{\rho}>0$ yields
\begin{equation} \frac{\cwinprod{W^\prime}{\rho}}{\cwinprod{W}{\rho}} \leq \left| \frac{\cwinprod{W^\prime}{\sigma}}{\cwinprod{W}{\sigma}} \right|. \end{equation}
Taking the infimum with respect to $\boldsymbol{\sigma} \in D_{\boldsymbol{W}}$ on the right hand side of this expression then yields the required result.

(iv) $\lambda \geq 1$ follows immediately from (ii). Clearly if $\boldsymbol{W}=\boldsymbol{W^\prime}$ than $\lambda=1$. Conversely, if $\lambda=1$, than (i) and (iii) imply that $\cwinprod{W}{\rho} \geq \cwinprod{W^\prime}{\rho}$ for all $\boldsymbol{\rho} \in C$. Hence if $\rho$ is a state compatible with $\boldsymbol{\rho}$, then $\Tr \left[ (p(\boldsymbol{W})-p(\boldsymbol{W^\prime}))\rho \right] \geq 0$. Let $\rho_i = \oper{\psi_i}{\psi_i}$ for $i=1,\ldots,D$, with the set $\{ \ket{\psi_i} \}_{i=1}^{D}$ being an orthonormal basis for $\mathcal{H}$. Since $\Tr(p(\boldsymbol{W})) = \Tr(p(\boldsymbol{W^\prime}))$, and
\begin{equation} 0 = \Tr((p(\boldsymbol{W})-p(\boldsymbol{W^\prime})) = \sum_{i=1}^D \Tr \left[ (p(\boldsymbol{W})-p(\boldsymbol{W^\prime}))\rho_i \right] \geq 0,  \end{equation}
we must have $\Tr \left[ (p(\boldsymbol{W})-p(\boldsymbol{W^\prime}))\rho_i \right] = 0$ for all $i$. But the orthonormal basis was arbitrary, and this implies that for all $\boldsymbol{\rho} \in C$ compatible with a state $\rho$, we have 
\begin{equation} \cwinprod{W}{\rho}-\cwinprod{W^\prime}{\rho} = \Tr \left[ (p(\boldsymbol{W})-p(\boldsymbol{W^\prime}))\rho \right] = 0. \end{equation}
However, by Lemma \ref{cw_id}, for any $\rho \in \mathcal{O}$, there exists $x \in (0,1)$ such that $\boldsymbol{\rho}(x) = (1-x)\boldsymbol{I} + x\boldsymbol{\rho}$ is compatible with a state $\rho(x)$. Hence $\cwinprodarg{W}{\rho}{x} = \cwinprodarg{W^\prime}{\rho}{x}$, and hence $\cwinprod{W}{\rho} = \cwinprod{W^\prime}{\rho}$ i.e. $\boldsymbol{W}=\boldsymbol{W^\prime}$. \end{proof}

\begin{proof}[Proof of Theorem \ref{cw_opt_thm}] First suppose that $\boldsymbol{W} = (1-\epsilon)\boldsymbol{W^\prime} + \epsilon \boldsymbol{P}$, with $\cwinprod{P}{\rho} \geq 0$ for all $\boldsymbol{\rho} \in \mathcal{O}$. Then $\cwinprod{W^\prime}{\rho} = \frac{1}{1-\epsilon} \left( \cwinprod{W}{\rho} - \epsilon \cwinprod{P}{\rho} \right)$, and hence if $\cwinprod{W}{\rho} <0$, then $\cwinprod{W^\prime}{\rho}<0$ also.

Conversely, take $\lambda$ defined as above. Then Lemma \ref{cw_lam}(iv) tells us that $\lambda \geq 1$. If $\lambda = 1$ then $\boldsymbol{W}=\boldsymbol{W^\prime}$ i.e. the theorem statement is satisfied with $\boldsymbol{P}=0$. If $\lambda > 1$, then define
\begin{equation} \boldsymbol{P} = \frac{1}{1-\lambda}(\lambda \boldsymbol{W} - \boldsymbol{W^\prime}); \quad \epsilon = 1- \frac{1}{\lambda}. \end{equation}
Then by Lemma \ref{cw_lam}(i)-(iii) above, $\cwinprod{P}{\rho} \geq 0$ for all $\boldsymbol{\rho} \in \mathcal{O}$, and by definition, $\boldsymbol{W} = (1-\epsilon)\boldsymbol{W^\prime} + \epsilon \boldsymbol{P}$ as required. \end{proof}

An example of a witness $\boldsymbol{P}$ with the above property is one where each $P_i$ is positive. But there also exist witnesses $\boldsymbol{P^\prime}$ where $P^\prime_i \ngeq 0$ (since e.g. we can add $\alpha \boldsymbol{1}$ to $P_i$ and subtract $\alpha \boldsymbol{1}$ from $P_j, j \neq i$ to form $\boldsymbol{P^\prime}$ such that $\cwinprod{P^\prime}{\rho} = \cwinprod{P}{\rho} \geq 0$ for all $\boldsymbol{\rho} \in C$. As a result, checking the condition of Theorem \ref{cw_opt_thm} is difficult, but such witnesses $\boldsymbol{P}$ can easily be used to `optimise' compatibility witnesses as outlined above.

To conclude this section, we note that there are many more results on optimal entanglement witnesses in \cite{LKCH00} than we have presented here. However, the difference in structure in compatibility and entanglement witnesses makes it difficult to imitate the ideas in \cite{LKCH00}.

\section{Semidefinite programming and the compatibility problem}

In this section we outline how the compatibility problem can be cast in the form of a semidefinite program, which makes it amenable to solution by numerical means. Semidefinite programming has been used in a number of ways to attack the separability problem \cite{DPS02,DPS,EHOC04,BrV04a,BrV04b}, and the ideas utilised in those papers (particularly those in \cite{DPS}) act as an inspiration for the application here.

We will start by giving a brief and by no means comprehensive introduction to semidefinite programming. Interested readers can refer to \cite{BL96, BLBk} for more details. We will then outline how semidefinite programming can be used to attack the compatibility problem, and we also illustrate how it be used to obtain some interesting analytical results, most notably to disprove the conjecture in \cite{BSS}.

\subsection{A summary of semidefinite programming}

A \emph{semidefinite program} (SDP) is a minimisation problem, which involves the minimisation of a linear function of a variable $\mathbf{x} \in \mathbb{R}^m$ subject to a linear matrix inequality. Formally, the problem is 
\begin{equation} \label{sdp_def}
\begin{array}{rl}
\textrm{Minimise} & \mathbf{c.x} \\
\textrm{subject to} & F(\mathbf{x}) \geq 0 
\end{array}
\end{equation}
where
\begin{equation} F(\mathbf{x}) = F_0 + \sum_{k=1}^m x_kF_k, \end{equation}
and $F_0, F_1, \ldots, F_m$ are fixed Hermitian matrices. The minimisation is performed over $\mathbf{x} \in \mathbb{R}^m$. The set of $\mathbf{x}$ for which the inequality $F(\mathbf{x}) \geq 0 $ holds is known as the \emph{feasible region}. The SDP is termed \emph{feasible} if this feasible region is non-empty. 

A semidefinite program is an example of a \emph{convex optimisation} problem, since the feasible region is convex i.e. if $F(\mathbf{x}), F(\mathbf{y}) \geq 0$, then for $\lambda \in [0,1]$,
\begin{equation} F(\lambda \mathbf{x} + (1-\lambda) \mathbf{y}) = \lambda F(\mathbf{x}) + (1-\lambda) F(\mathbf{y}) \geq 0. \end{equation}
It is this convexity that has allowed fast algorithms to be developed to solve semidefinite programs numerically, and details about these algorithms can be found in \cite{BL96}.

An important structural feature of semidefinite programming is a feature known as \emph{duality}. Associated with a problem of the form (\ref{sdp_def}), usually known as the primal problem, is a so called dual problem:
\begin{equation} \label{sdp_dual}
\begin{array}{rl}
\textrm{Maximise} & -\Tr(F_0Z) \\
\textrm{subject to} & Z=Z^\dagger, Z \geq 0, \\
& \Tr(F_iZ) = c_i. 
\end{array}
\end{equation}
This dual problem is a semidefinite program as it can be cast in the form of (\ref{sdp_def}) \cite{BL96}. There are a number of relationships between these two problems.
\begin{itemize}
\item Let $\mathbf{x}, Z$ be feasible solutions to the primal and dual problem respectively. Then
\begin{equation} \mathbf{c.x} + \Tr(F_0Z) = \Tr(F(\mathbf{x})Z) \geq 0 \label{sdp_wd} \end{equation}
and so $-\Tr(F_0Z) \leq \mathbf{c.x}$ i.e. any feasible $\mathbf{x}$ can be used to compute an upper bound on $-\Tr(F_0Z)$, and any feasible $Z$ can be used to compute a lower bound on $\mathbf{c.x}$. 
\item We say the primal problem is \emph{strictly feasible} if there exits $\mathbf{x}$ such that $F(\mathbf{x}) > 0$ (i.e. a strict inequality is satisfied). Similarly, the dual problem is strictly feasible if there exists $Z > 0 $ satisfying $\Tr(F_iZ) = c_i$. If \emph{either} of the two problems are strictly feasible, then the optimium values for both problems are equal \cite{BL96}.
\item Finally, if $\mathbf{x},Z$ achieve the optimum values for the primal and dual problems respectively, then (\ref{sdp_wd}) implies that $\Tr(F(\mathbf{x})Z) = 0$. However, since $F(\mathbf{x}) \geq 0$ and $Z \geq 0$, $F(\mathbf{x})Z = 0$. This is known as the \emph{complementary slackness} condition.
\end{itemize}

\subsection{Formulating the compatibility problem as an SDP}

We will now discuss how to express the compatibility problem as a semidefinite program of the form (\ref{sdp_def}). The idea is this: $F(\mathbf{x})$ will represent the potential compatible state, and the inequality $F(\mathbf{x}) \geq 0$ will correspond to the necessary positivity of any state compatible with the $(n-1)$-party states. Within the definition of $F(\mathbf{x})$, the matrix $F_0$ will represent the part of any compatible state that is fixed by the reduced states. The other matrices represent the freedom in the potential choice of any compatible state, and as a result do not change the partial traces of the potential state.

Let $B_\mathbf{m}$ be the basis elements for $\mathcal{B}(\mathcal{H})$ as defined in (\ref{basis}). Suppose $\rho \in \mathcal{B}(\mathcal{H})$ is a quantum state, with $\Tr_k (\rho) = \rho^{(k)}$. Any $\rho$ can be written in the form
\begin{equation} \rho = \sum_{\mathbf{m}} \rho_{\mathbf{m}} B_\mathbf{m}. \end{equation}
However, defining $I_k = \{ \mathbf{m} \ | \ m_k=1 \}$, and, for $\mathbf{m} \in I_k$, 
\begin{equation} B^{(k)}_{\mathbf{m}} = \otimes_{i \neq k} B_{i,m_i} \end{equation}
then we can observe that
\begin{equation} \Tr_k(\rho) = \sum_{\mathbf{m} \in I_k} \rho_\mathbf{m} B^{(k)}_{\mathbf{m}} \end{equation}
i.e. the condition $\Tr_k (\rho) = \rho^{(k)}$ fixes $\rho_\mathbf{m}$ for $\mathbf{m} \in I_k$. Since $\cup_{k=1}^n I_k = I_C$ (see (\ref{is})), the $n$ partial trace conditions fix $\rho_\mathbf{m}$ for $\mathbf{m} \in I_C$ \footnote{The sets $I_k$ are not disjoint, and so it would appear that we are placing multiple conditions on $\rho_{\mathbf{m}}$; however, the actual conditions are equal when overlapping trace conditions like those in (\ref{trace}) are imposed on $\rho^{(1)}, \ldots, \rho^{(n)}$.}. With this in mind, we define 
\begin{equation} B_0 = \sum_{\mathbf{m} \in I_C} \rho_{\mathbf{m}} B_\mathbf{m} \label{b0} \end{equation} 
where the coefficients $\rho_{\mathbf{m}}$ are fixed by the appropriate partial trace conditions.

The remaining basis elements ($B_\mathbf{m}$, where $\mathbf{m} \in I \setminus I_C$) are a tensor product of traceless matrices (since if $\mathbf{m} \notin I_C, m_k \neq 1$ for all k, and $\Tr(B_{k,m}) = 0$ when $m \neq 1$), and so yield zero under tracing of any system. As a result, the operator
\begin{equation}
B(\mathbf{x}) = B_0 + \sum_{\mathbf{m} \in I / I_C} x_{\mathbf{m}} B_\mathbf{m}
\end{equation}
(where $\mathbf{x} \in \mathbb{R}^{(|I|-|I_C|)}$) also satisfies the appropriate partial trace conditions. Therefore, the compatibility problem has now been reduced to the problem of finding $\mathbf{x}$ such that $B(\mathbf{x}) \geq 0$.

An equivalent problem is to consider the SDP given by
\begin{equation} \label{sdp_com}
\begin{array}{rl}
\textrm{Minimise} & t \\
\textrm{subject to} & B_0 + \sum_{\mathbf{m} \in I / I_C} x_{\mathbf{m}} B_\mathbf{m} + t\boldsymbol{1} \geq 0. 
\end{array}
\end{equation}
The effect of this problem is to find an operator with the appropriate reduced states with the maximum minimal eigenvalue (given by $-t$). If $t>0$, then $B(\mathbf{x})$ can never be positive, and the $(n-1)$-party states are not compatible with a quantum state of the entire system.

Before we continue, we note that there is a second method of calculating the fixed operator $B_0$: we note that the expression for $B_0$ given in (\ref{b0}) is similar to that given in (\ref{cw_basis}), and so the techniques used in the proof of Proposition \ref{prop_cwtest} can be used to deduce that
\begin{equation} B_0 = -\sum_{A \subseteq N, A \neq \emptyset} \frac{(-1)^{|A|}}{d_A} \rho_{N \setminus A} \otimes 1_A, \end{equation}
where $\rho_B$ $(B \subset N)$ is the subsystem state for the systems labelled by the elements of $B$ obtained by partial traces of appropriate $(n-1)$-party states.

\subsection{The dual problem and compatibility witnesses}

The dual problem for the compatibility problem is given by
\begin{equation} 
\begin{array}{rl}
\textrm{Maximise} & -\Tr(B_0Z) \\
\textrm{subject to} & Z=Z^\dagger, Z \geq 0, \\
& \Tr(Z) = 1, \\
& \Tr(B_\mathbf{m}Z) = 0 \textrm{ for all } \mathbf{m} \in I \setminus I_C. 
\end{array}
\end{equation}

However, $Z=p(\boldsymbol{W})$ for some compatibility witness $\boldsymbol{W}$: Since $I \setminus I_C = \{ \mathbf{m} \ | \ m_k = 2, \ldots, d_k^2, k=1,\ldots,n \}$, and $B_{k,2}, \ldots, B_{k,d_k^2}$ forms a basis for the space of traceless operators on $\mathcal{H}_k$, Proposition 5 implies that $Z=p(\boldsymbol{W})$ for some $\boldsymbol{W} \in \mathcal{W}$. 

The upshot of this is that if the optimum value of the dual problem is positive, the $Z$ which realises this maximum is such that $Z=p(\boldsymbol{W})$, and 
\begin{eqnarray}
0 > \Tr(B_0Z) &=& \Tr \left[B_0 \left( \sum_{k=1}^n W_i \otimes \boldsymbol{1}_i \right) \right] \\
&=& \sum_{k=1}^n \Tr(\rho_i W_i) \\
&=& \cwinprod{W}{\rho}
\end{eqnarray}
i.e. $\boldsymbol{W}$ is a compatibility witness that detects the fact that the $(n-1)$-party reduced states $\rho_i$ are not compatible with a quantum state of the overall state \footnote{This dual problem is strictly feasible (for example, choose $Z=\boldsymbol{1}/(\prod_i d_i)$ to satisfy $Z>0$ and the constraints of the problem), so if the minimum of the primal problem is $t>0$, then there exists $Z$ satisfying the constraints of the problem such that $\Tr(B_0Z)=-t<0$ i.e. a compatibility witness can always be found.}.

\subsection{Complexity}

The SDP (\ref{sdp_def}) involves optimising over the space $\mathbb{R}^m$, and dealing with the positivity of $N \times N$ Hermitian matrices. Numerical SDP solvers are known to have complexity $O(m^2N^{5/2})$ ($O(N^{1/2})$ iterations, each taking time $O(m^2N^2)$ \cite{BL96}) \footnote{Note that in all our complexity calculations, we assume that all numerical values are held to a fixed precision, and so all individual calculations are of time $O(1)$.}.

In our case, if we define $D = \prod_i d_i$, then
\begin{itemize}
\item The number of variables is equal to $|I| - |I_C| +1 = \prod_{i=1}^n (d_i^2-1) + 1 = O(D^2)$;
\item The dimension of the matrices equals $D$.
\end{itemize}
and so the complexity of the compatibility SDP is $O(D^{13/2})$. This is polynomial in the dimension of every individual system, but is exponential in the number of parties i.e. if $d_i=d$, then $D=d^n$, leading to exponential complexity in $n$. However, the complexity of the algorithm is still polynomial in the problem size, since specifying the reduced state which excludes system $k$ involves specifying a $D/d_k \times D/d_k$ Hermitian matrix i.e. $(D/d_k)^2$ parameters. So in total we have a problem size of 
\begin{equation} s=O\left(D^2 \sum_{k=1}^n d_k^{-2} \right)=O(D^2) \end{equation}
hence the time complexity is bounded above by a polynomial in $s$. Hence the problem lies in the complexity class $P$. The exponential complexity in $n$ merely reflects the fact that we must supply an exponentially increasing amount of initial data in the form of the reduced states.

Note that this does not contradict the results of \cite{Liu06a} discussed earlier, as the compatibility problem considered there involves having only a number of reduced states that is polynomial in the number of quantum systems $n$, and each reduced state describes the state of a number of systems below some fixed constant. Hence the problem size here is only a polynomial in $n$, leading to very different conclusions regarding the complexity of the problem. We will return to this issue later.

\subsection{New insights from the compatibility SDP}

We have produced computer code for the compatibility SDP using MATLAB and a freely available SDP numerical solver known as SEDUMI \cite{SEDUMI}.
In much the same way as was achieved when semidefinite programming was applied to the separability problem, analytical results can be gained by inspection of numerical results due to the accuracy numerical SDP solvers work to.

A simple example of this comes from considering the following two-qubit subsystem states of a three party system: 
\begin{equation} \rho^{(1)} = \rho^{(2)} = \rho^{(3)} = \oper{\Psi_+}{\Psi_+}; \quad \ket{\Psi_+} = \frac{1}{\sqrt{2}} \left( \ket{00} + \ket{11} \right) \end{equation}
which are not compatible with an overall state (if qubits 1,2 are in a pure entangled state, then qubit 3 must be in a pure state, which is not the case here). The primal SDP confirms this incompatibility, and the dual SDP can be used to determine a numerical compatibility witness. From this result, an analytic compatibility witness can be derived that detects this incompatibility. Let
\begin{equation} Z = \frac{1}{2} \left( \oper{\psi_1}{\psi_1} + \oper{\psi_2}{\psi_2} \right) \end{equation}
where
\begin{eqnarray}
\ket{\psi_1} &=& \frac{1}{2} \left( \ket{000} - \ket{011} - \ket{101} - \ket{110} \right), \\
\ket{\psi_2} &=& \frac{1}{2} \left( \ket{001} + \ket{010} + \ket{100} - \ket{111} \right).
\end{eqnarray}
This can easily be checked to be a compatibility witness by using Proposition \ref{prop_cwtest}. Due to the symmetry of $Z$, all the three two-party reduced states of $Z$ are equal:
\begin{equation}
Z_{12} = Z_{13} = Z_{23} = \frac{1}{2} \left( \oper{\Psi_-}{\Psi_-} + \oper{\Phi_+}{\Phi_+} \right);
\end{equation}
\begin{equation}
\ket{\Psi_-} = \frac{1}{\sqrt{2}} \left( \ket{00} - \ket{11} \right), \ket{\Phi_+} = \frac{1}{\sqrt{2}} \left( \ket{01} + \ket{10} \right),
\end{equation}
and all the one-party reduced states of $Z$ are equal to $\boldsymbol{1}/2$. Hence a compatibility witness $\boldsymbol{W}$ corresponding to $Z$ is given by
\begin{equation} W_1 = W_2 = W_3 = -\frac{1}{12} + \frac{1}{4} \left( \oper{\Psi_-}{\Psi_-} + \oper{\Phi_+}{\Phi_+} \right) \end{equation}
which leads to $\sum_{i=1}^3 \Tr(\rho^{(i)}W_i) = -1/4$, detecting the incompatibility as required.

The compatibility SDP can be used to help us produce a set of 2-qubit reduced states that disproves the conjecture in \cite{BSS}. For the case of 3 qubits, the conjecture states that $\rho_{12}, \rho_{13}, \rho_{23}$ are the 2-qubit reduced states of a three-qubit state $\rho$ if and only if 
\begin{equation} 0 \leq \Delta(\boldsymbol{\rho}) \equiv \boldsymbol{1} - \rho_1 - \rho_2 - \rho_3 + \rho_{12} + \rho_{13} + \rho_{23} \leq \boldsymbol{1} \end{equation}
where the reduced states are padded out with identities as stated in Theorem \ref{com_conj}, and $\boldsymbol{\rho}$ is the 3-vector of 2-party reduced states.

Let $\sigma = \frac{1}{4} \left( \oper{1}{1} \otimes \boldsymbol{1} \otimes \boldsymbol{1} \right)$. Let us define the following two-qubit subsystem states of a three qubit system:
\begin{equation}
\rho^{(k)}(p) = (1-p)\Tr_k \sigma + p\oper{\Psi_+}{\Psi_+} 
\end{equation}
where $p \in [0,1]$. This set of subsystem states is clearly compatible with an overall state when $p=0$, and incompatible with any overall state when $p=1$. We can calculate $\Delta(\boldsymbol{\rho})$ for these reduced states to be
\begin{eqnarray}
\Delta(\sigma) &=& (1-p)\boldsymbol{1}/4 + 3p/2\left( \oper{\phi_1}{\phi_1} + \oper{\phi_2}{\phi_2} \right) \nonumber \\
&-& p/2 \left( \oper{\psi_1}{\psi_1} + \oper{\psi_2}{\psi_2} \right) \end{eqnarray}
where
\begin{eqnarray}
\ket{\phi_1} &=& \frac{1}{\sqrt{12}} \left( 3\ket{000} + \ket{011} + \ket{101} + \ket{110} \right), \\
\ket{\phi_2} &=& \frac{1}{\sqrt{12}} \left( \ket{001} + \ket{010} + \ket{100} + 3\ket{111} \right),
\end{eqnarray}
and hence $0 \leq \bra{\psi}\Delta\ket{\psi} \leq 1$ when $p \leq 1/3$. If the conjecture is true, then when $p \leq 1/3$, there should be an overall state compatible with the above subsystem states. However, the primal SDP can be used to show this is not true, and produce a numerical compatibility witness illustrating this fact. From this, an analytical compatibility witness can be obtained. For the case $p=1/4$, it comes from the operator 
\begin{widetext}
\begin{equation}
Z = \frac{1}{100000}\left(
\begin{array}{cccccccc}
  2959&     0&     0&  -102&     0& -1715& -1715&     0 \\
     0& 24865&  1005&     0&   766&     0&     0& -1715 \\
     0&  1005& 24865&     0&   766&     0&     0& -1715 \\
  -102&     0&     0& 45033&     0&   766&   766&     0 \\
     0&   766&   766&     0&    46&     0&     0&  -102 \\
 -1715&     0&     0&   766&     0&  1006&  1005&     0 \\
 -1715&     0&     0&   766&     0&  1005&  1006&     0 \\
     0& -1715& -1715&     0&  -102&     0&     0&   228 
\end{array} \right)
\end{equation}
\end{widetext}
which can be verified to be positive, to be of the form $Z=p(\boldsymbol{W})$ through Proposition 6, and to witness the incompatibility of the above subsystem states for $p=1/4$. This example hence disproves the conjecture given in \cite{BSS}. This witness can also be seen to be an example of the optimisation process we saw in section \ref{ss_ocw}: Defining 
\begin{equation} Z^\prime = 10^{-6}\textrm{diag}(0,0,0,0,2913,23859,23859,44805) + \alpha \boldsymbol{1} \end{equation} 
and considering the decomposition $Z = (Z+Z^\prime) - Z^\prime$, then it can be shown that
\begin{itemize}
\item[(a)] $Z+Z^\prime = \Delta(Z+Z^\prime)$;
\item[(b)] $Z^\prime = p(\boldsymbol{P})$;
\item[(c)] For large enough $\alpha$, $\cwinprod{P}{\rho} \geq 0$ for all $\boldsymbol{\rho} \in \mathcal{O}$. 
\end{itemize}
i.e. $Z$ is an optimisation of $\Delta(Z+Z^\prime)$, and can detect incompatibility that the condition $\Delta(\boldsymbol{\rho}) \geq 0$ cannot.

\section{Solutions of variants of the compatibility problem}

This new technique of using semidefinite programming to numerically solve the compatibility problem can be applied to important variants of the compatibility problem we have considered. We discuss these applications in this section.

\subsection{Partial knowledge of reduced states} \label{ss_par_know}

A simple modification of the problem is to consider a situation where we are only given a partial subset of the reduced states. The problem (\ref{sdp_com}) can easily be modified to solve this problem: The number of basis operators $B_\mathbf{m}$ that appear in the fixed operator $B_0$ is reduced according to which reduced states are known, and the coefficients of the remaining basis operators are then determined by the SDP solver. The dual problem to this program will still produce a compatibility witness $Z=p(\boldsymbol{W})$, but with some extra conditions of the form $\Tr(ZB_\mathbf{m})=0$ for the the basis operators that are no longer fixed in the operator $B_0$.

The complexity of the problem depends on the problem size i.e. the number of known reduced states, and the number of systems described by these states. Let us consider the case discussed in \cite{Liu06a} (to reiterate: we have a number of reduced states $p(n)$, a polynomial in $n$, each describing a number of systems less than some constant $k$). In this case, the problem size is $O(p(n))$, whereas the dimension of the matrices involved is $D$, and there are $O(D^2)$ free variables (since fixing polynomially many reduced states describing at most $k$ systems can only fix polynomially many coefficients in $B_0$). Hence the complexity is again $O(D^{13/2})$, but this is no longer polynomial in the problem size. 

This is simply one complexity scenario we could imagine; we could place other conditions on the number of reduced states we know, and how many states they can describe. As complexity is not our main focus here, we leave these questions for further investigation.

\subsection{The bosonic and fermionic compatibility problems}

If a density state $\rho$ represents the state of either $n$ bosons or $n$ fermions, there are some additional constraints that much be placed on $\rho$. A pure state $\ket{\psi} \in \mathcal{H}$ representing $n$ bosons must be totally \emph{symmetric} under particle exchanges (or swaps between Hilbert spaces); whereas any pure state of $n$ fermions should be totally \emph{antisymmetric} under particle exchanges. 

Since we are dealing with identical particles, the $n$ single particle Hilbert spaces are all identical; let their dimension be $d$. Also, let $B_{k,m}$ be equal for all $k$ to $B_m$.  The total (anti)symmetry means that all of the $k$-party reduced states will be identical; as a result we will refer to any $k$-party reduced state as \emph{the} $k$-particle reduced state.

The space of totally symmetric pure states has an orthonormal basis $\{ \ket{S_{\textbf{x}}} \}_{1 \leq x_1 \leq \ldots \leq x_n \leq d}$, where
\begin{equation}
\ket{S_{\textbf{x}}} = \left( \frac{n_1({\textbf{x}})!\ldots n_d({\textbf{x}})!}{n!} \right)^{1/2} \sum_{\rho \in S_n}\ket{x_{\rho(1)}}\ldots\ket{x_{\rho(n)}} \end{equation}
(where $n_k(\textbf{x})$ is defined to be the number of instances of $k$ within $\textbf{x}$). The projection on this subspace is given by
\begin{equation} P_S = \sum_{x_1 \leq \ldots \leq x_n} \oper{S_{\textbf{x}}}{S_{\textbf{x}}} \end{equation} A density state $\rho$ is then totally symmetric if and only if $P_S\rho P_S = \rho$. However, since $\rho$ is positive and of trace one, this relation is satisfied if and only if $\Tr(P_S\rho P_S) = \Tr(\rho) = 1$. Finally, since $P_S^2 = P_S$, and by the cyclic property of the trace function, we have that $\rho$ is totally symmetric if and only if $\Tr(\rho P_S) = 1$. Noting that $\Tr(P_S\rho P_S) \leq \Tr(\rho) = 1$ (because $P_S$ is a projection), we can change the condition to $\Tr(\rho P_S) \geq 1$. Taking all this into account, we can hence modify problem (\ref{sdp_com}) to accommodate the totally symmetric condition:
\begin{equation} \label{sdp_com_boson}
\begin{array}{rl}
\textrm{Minimise} & t \\
\textrm{subject to} & B_0 + \sum_{\mathbf{m} \in I \setminus I_C} x_{\mathbf{m}} B_\mathbf{m} + t\boldsymbol{1} \geq 0 \\
& -1 + \Tr(B_0 P_S) + \sum_{\mathbf{m} \in I \setminus I_C} x_{\mathbf{m}} \Tr(B_\mathbf{m} P_S) \geq 0 
\end{array}
\end{equation} 
(note that two linear matrix inequalities $B_1(\mathbf{x}) \geq 0, B_2(\mathbf{x}) \geq 0$ can be combined into one of the form $B_1(\mathbf{x}) \oplus B_2(\mathbf{x}) \geq 0$, so this really is a semidefinite program). 

We can produce a similar SDP for the fermionic case. The space of totally antisymmetric pure states is spanned by an orthonormal basis $\{ \ket{A_{\textbf{x}}} \}_{1 \leq x_1 < \ldots < x_n \leq d}$, where
\begin{equation}
\ket{A_{\textbf{x}}} = \sqrt{ \frac{1}{n!} } \sum_{\rho \in S_n} \mathcal{E}(\rho) \ket{x_{\rho(1)}}\ldots\ket{x_{\rho(n)}}
\end{equation}
where $\mathcal{E}(\rho)$ denotes the sign (parity) of the permutation $\rho \in S_n$. The projection onto the antisymmetric subspace is hence given by
\begin{equation} P_A = \sum_{x_1 < \ldots < x_n} \oper{A_{\textbf{x}}}{A_{\textbf{x}}} \end{equation}
and similarly the semidefinite program for the fermionic compatibility problem is given by
\begin{equation} \label{sdp_com_fermion}
\begin{array}{rl}
\textrm{Minimise} & t \\
\textrm{subject to} & B_0 + \sum_{\mathbf{m} \in I \setminus I_C} x_{\mathbf{m}} B_\mathbf{m} + t\boldsymbol{1} \geq 0 \\
& -1 + \Tr(B_0 P_A) + \sum_{\mathbf{m} \in I \setminus I_C} x_{\mathbf{m}} \Tr(B_\mathbf{m} P_A) \geq 0 
\end{array}
\end{equation} 

We can reduce number of free variables in both of these problems by noticing that for both the bosonic and fermionic case, the density matrix of $n$ particles is symmetric under any particle exchange (in the antisymmetric case, $\ket{A_\mathbf{x}}$ is totally antisymmetric, but any operator $\oper{A_\mathbf{x}}{A_\mathbf{y}}$ is symmetric because any factors of $-1$ cancel out). Hence the density matrices only depend on symmetrised basis operators $\{ B^S_\mathbf{m} \}_{1 \leq m_1 \leq \ldots \leq m_n \leq d}$, where
\begin{equation} B_\mathbf{m}^S = \sum_{\rho \in S_n} \left( \bigotimes_{k=1}^n B_{m_{\rho(k)}} \right). \end{equation}
Further to this, the two semidefinite programs can be modified again to deal with cases where we have only partial knowledge of the subsystem states, by adjusting the inequalities in the same way as outlined in section \ref{ss_par_know}. This would allow us for example to deal with the $N$-representability problem discussed in section \ref{ss_prev_res}.

\subsection{Determination of ground state energies of identical particle systems}

We can further modify the semidefinite programs (\ref{sdp_com_boson}) and (\ref{sdp_com_fermion}) to create a program for determining the ground state energy of a system of $n$ identical particles with a Hamiltonian that consists only of two-body interactions. The (anti)symmetry of any state of these particles implies that we can replace the $n$-body Hamiltonian with an effective two-body Hamiltonian $H^{(2)}$. The 2-particle reduced state $\rho^{(2)}$ of a state $\rho$ can then be used to evaluate the energy $E=\Tr(H^{(2)}\rho^{(2)})$ (see e.g. \cite{Maz06} for more details).  

We can use semidefinite programming to minimise $E$, subject to constraints that ensures $\rho^{(2)}$ is a valid 2-particle density state. The following program calculates $E$ for a set of $n$ bosons:

\begin{equation} \label{sdp_com_gse}
\begin{array}{rl}
\textrm{Minimise} & \Tr(H^{(2)}\rho^{(2)}) \\
\textrm{subject to} & \rho^{(2)} = \Tr_{3, \ldots, N} \rho \\
& \rho \geq 0; \Tr(\rho P_S) \geq 1
\end{array}
\end{equation}
(by writing $\rho = \sum_{1 \leq m_1 \leq \ldots \leq m_n \leq d} x_{\mathbf{m}} B^S_\mathbf{m}$, the above problem can be cast in a semidefinite program of the form of (\ref{sdp_def})). For a set of $n$ fermions, we simply replace $P_S$ by $P_A$. However, the complexity of this problem will again be exponential in $n$, and so for large numbers of particle this solution will be inefficient. Indeed, recent results showing that the $N$-representability problem is QMA-complete \cite{LCV06} illustrates that an efficient solution to this problem (one with running time polynomial in $n$) is highly unlikely to exist.

\subsection{Other variants}

The key fact that allows us to apply the ideas of semidefinite programming to these variants of the subsystem problem is the fact that both the set of (anti)symmetric states and the set of any fixed subset of the subsystem states compatible with an overall state are both \emph{convex} sets. A variant of this problem that would be difficult to solve using this algorithm would be the problem where we wish to find a compatible state with a particular fixed spectrum, because the set of operators with a fixed spectrum is not convex. A different method would have to be used in this case. 

\section{Further ideas}

In this paper we have illustrated how we can numerically solve the compatibility problem using semidefinite programming, and the fast algorithms available to solve semidefinite programs make such a task operationally feasible \footnote{On a 2.8GHz Pentium 4 with 512Mb of memory, the problem is solved in less than one second for three qubits.}. With only slight modifications, the SDP can deal with a situation where we only have partial knowledge of the proper subsystem states, or where we are dealing with a set of identical bosonic or fermionic particles.

The introduction of \emph{compatibility witnesses} in this paper marks a move away from the usual methods used to attack the compatibility problem, and as a result there is possibly more that can be said about the problem. We will discuss some of those possibilities here.

Having disproved the conjecture from \cite{BSS}, we are left then with what further necessary conditions we can find for the compatibility problem, and what (finite or infinite) subset of these form a sufficient condition for the problem. One possible source of these could come from a study of positive maps: We can view the operator $\Delta$ as a positive map on the space $\mathcal{B}(\mathcal{H})$ that only depends on the reduced states $\rho_A$. So, finding other positive maps only depending on the reduced states could give us more necessary conditions. Finding whether these conditions could be sufficient however would be difficult.

Another possible source of ideas is a study of the semidefinite program itself. The problem of minimising the maximum eigenvalue of a symmetric matrix that is an argument of an affine parameter has been well studied (see references in \cite{BL96}), and these ideas could lead to new insights about this problem.

One oddity about this problem is the fact that many of the useful results obtained so far only apply when we have an odd number of systems. We can obtain conditions for the even case by treating two individual systems as one system, but this obviously does not reflect the full generality of the problem. Despite some effort, we have not been able to make any further progress in this case.

Finally, we presented a number of properties of compatibility witnesses, but we have not given a complete characterisation of these witnesses, and knowing more about their structure would be very helpful indeed. However, we will make a further observation here. Let us continue to explore the case where we have an odd number of systems ($n$ is odd). For $Z \in \mathcal{B}(\mathcal{H})$, let
\begin{equation} \Delta(Z) = \sum_{A \subset N} (-1)^{|A|} Z_A \otimes \boldsymbol{1}_{N \setminus A} \end{equation}
(This is equal to the expression defined in Theorem \ref{com_conj} when an overall state exists). Then if $Z\geq0$, $\Delta(Z)=p(\boldsymbol{W})$: Let $T=T_1 \otimes \ldots \otimes T_n$, with $T_i \in \mathcal{B}(\mathcal{H}_i), \Tr(T_i)=0$. Then
\begin{equation} \Tr(T\Delta(Z)) = \sum_{A \subset N} (-1)^{|A|} \Tr(Z_A T_A) \Tr(T_{N \setminus A}) = 0 \end{equation}
and since $\Delta$ is a positive map \cite{Hall}, Proposition \ref{prop_t2} imples that $\Delta(Z)=p(\boldsymbol{W})$. However, this is does not form the entire set of compatibility witnesses: For $\boldsymbol{\rho} \notin C$, but $\Delta(\boldsymbol{\rho}) \geq 0$, then, for all $Z \geq 0$,
\begin{eqnarray}
\Tr(B_0 \Delta(Z)) &=& \Tr(\Delta(B_0) Z) \\
									 &=& \Tr(\Delta(\boldsymbol{\rho})Z) \geq 0
\end{eqnarray}
where the first line follows from the fact the adjoint map of $\Delta$ is itself, and the second line follows since the partial traces of $B_0$ coincide with the subsystem states in $\boldsymbol{\rho}$. All of this implies the following theorem:

\begin{thm} If $\boldsymbol{\rho} \notin C$, but $\Delta(\boldsymbol{\rho}) \geq 0$, then for all compatibility witnesses $\boldsymbol{W}$ satisfying $p(\boldsymbol{W}) = \Delta(Z)$ for some $Z \geq 0$,  $\cwinprod{W}{\rho} \geq 0$. \end{thm}

Again using our analogy with the separability problem, we can think of sets of subsystem states that are incompatible with a full system state, but satisfy $\Delta(\boldsymbol{\rho}) \geq 0$ as analogous to entangled states with a positive partial transpose, in the sense that it is difficult to explicitly find entanglement witnesses that detect these states. The compatibility witness given at the end of the last section is an example of a witness that does detect incompatible subsystems states satisfying $\Delta(\boldsymbol{\rho}) \geq 0$, and furthermore it comes from an enhancement of a witness of the form $\Delta(Z)$. 

There is great potential for powerful compatibility witnesses to be constructed using this witness optimisation technique. Exploring this avenue further and looking for other methods to construct compatibility witnesses not of the form $\Delta(Z)$ could well be a good step into producing new necessary criteria for the problem.

\acknowledgments{The author would like to thank Tony Sudbery for reading this manuscript and his continued support, and the Engineering and Physical Sciences Research Council (U.K.) for funding this research.}

\bibliography{ccrefs}

\begin{thebibliography}{31}
\expandafter\ifx\csname natexlab\endcsname\relax\def\natexlab#1{#1}\fi
\expandafter\ifx\csname bibnamefont\endcsname\relax
  \def\bibnamefont#1{#1}\fi
\expandafter\ifx\csname bibfnamefont\endcsname\relax
  \def\bibfnamefont#1{#1}\fi
\expandafter\ifx\csname citenamefont\endcsname\relax
  \def\citenamefont#1{#1}\fi
\expandafter\ifx\csname url\endcsname\relax
  \def\url#1{\texttt{#1}}\fi
\expandafter\ifx\csname urlprefix\endcsname\relax\def\urlprefix{URL }\fi
\providecommand{\bibinfo}[2]{#2}
\providecommand{\eprint}[2][]{\url{#2}}

\bibitem[{\citenamefont{Coffman et~al.}(2000)\citenamefont{Coffman, Kundu, and
  Wootters}}]{CKW00}
\bibinfo{author}{\bibfnamefont{V.}~\bibnamefont{Coffman}},
  \bibinfo{author}{\bibfnamefont{J.}~\bibnamefont{Kundu}}, \bibnamefont{and}
  \bibinfo{author}{\bibfnamefont{W.~K.} \bibnamefont{Wootters}},
  \bibinfo{journal}{Phys. Rev. A} \textbf{\bibinfo{volume}{61}},
  \bibinfo{pages}{052306} (\bibinfo{year}{2000}), \eprint{quant-ph/9907047}.

\bibitem[{\citenamefont{Osborne and Verstraete}(2006)}]{OsV06}
\bibinfo{author}{\bibfnamefont{T.~J.} \bibnamefont{Osborne}} \bibnamefont{and}
  \bibinfo{author}{\bibfnamefont{F.}~\bibnamefont{Verstraete}},
  \bibinfo{journal}{Phys. Rev. Lett.} \textbf{\bibinfo{volume}{96}},
  \bibinfo{pages}{220503} (\bibinfo{year}{2006}), \eprint{quant-ph/0502176}.

\bibitem[{\citenamefont{Butterley et~al.}(2006)\citenamefont{Butterley,
  Sudbery, and Szulc}}]{BSS}
\bibinfo{author}{\bibfnamefont{P.}~\bibnamefont{Butterley}},
  \bibinfo{author}{\bibfnamefont{A.}~\bibnamefont{Sudbery}}, \bibnamefont{and}
  \bibinfo{author}{\bibfnamefont{J.}~\bibnamefont{Szulc}},
  \bibinfo{journal}{Found. Phys} \textbf{\bibinfo{volume}{36}},
  \bibinfo{pages}{83} (\bibinfo{year}{2006}), \eprint{quant-ph/0407227}.

\bibitem[{\citenamefont{Horodecki et~al.}(2001)\citenamefont{Horodecki,
  Horodecki, and Horodecki}}]{HHHBk}
\bibinfo{author}{\bibfnamefont{M.}~\bibnamefont{Horodecki}},
  \bibinfo{author}{\bibfnamefont{P.}~\bibnamefont{Horodecki}},
  \bibnamefont{and}
  \bibinfo{author}{\bibfnamefont{R.}~\bibnamefont{Horodecki}}, in
  \emph{\bibinfo{booktitle}{Quantum Information: An Introduction to Basic
  Theoretical Concepts and Experiments}}, edited by
  \bibinfo{editor}{\bibfnamefont{G.}~\bibnamefont{Alber}}
  (\bibinfo{publisher}{Springer-Verlag (Germany)}, \bibinfo{year}{2001}),
  chap.~\bibinfo{chapter}{5}, pp. \bibinfo{pages}{151--195},
  \eprint{quant-ph/0109124}.

\bibitem[{\citenamefont{Boyd and Vandenberghe}(1996)}]{BL96}
\bibinfo{author}{\bibfnamefont{S.}~\bibnamefont{Boyd}} \bibnamefont{and}
  \bibinfo{author}{\bibfnamefont{L.}~\bibnamefont{Vandenberghe}},
  \bibinfo{journal}{SIAM Review} \textbf{\bibinfo{volume}{38}},
  \bibinfo{pages}{49} (\bibinfo{year}{1996}).

\bibitem[{\citenamefont{Boyd and Vandenberghe}(2004)}]{BLBk}
\bibinfo{author}{\bibfnamefont{S.}~\bibnamefont{Boyd}} \bibnamefont{and}
  \bibinfo{author}{\bibfnamefont{L.}~\bibnamefont{Vandenberghe}},
  \emph{\bibinfo{title}{Convex Optimization}} (\bibinfo{publisher}{Cambridge
  University Press}, \bibinfo{year}{2004}).

\bibitem[{\citenamefont{Brand{\~a}o and Vianna}(2004{\natexlab{a}})}]{BrV04b}
\bibinfo{author}{\bibfnamefont{F.~G. S.~L.} \bibnamefont{Brand{\~a}o}}
  \bibnamefont{and} \bibinfo{author}{\bibfnamefont{R.~O.}
  \bibnamefont{Vianna}}, \bibinfo{journal}{Phys. Rev. A}
  \textbf{\bibinfo{volume}{70}}, \bibinfo{pages}{062309}
  (\bibinfo{year}{2004}{\natexlab{a}}), \eprint{quant-ph/0405008}.

\bibitem[{\citenamefont{Brand{\~a}o and Vianna}(2004{\natexlab{b}})}]{BrV04a}
\bibinfo{author}{\bibfnamefont{F.~G. S.~L.} \bibnamefont{Brand{\~a}o}}
  \bibnamefont{and} \bibinfo{author}{\bibfnamefont{R.~O.}
  \bibnamefont{Vianna}}, \bibinfo{journal}{Phys. Rev. Lett.}
  \textbf{\bibinfo{volume}{93}}, \bibinfo{pages}{220503}
  (\bibinfo{year}{2004}{\natexlab{b}}), \eprint{quant-ph/0405063}.

\bibitem[{\citenamefont{Doherty et~al.}(2002)\citenamefont{Doherty, Parrilo,
  and Spedalieri}}]{DPS02}
\bibinfo{author}{\bibfnamefont{A.~C.} \bibnamefont{Doherty}},
  \bibinfo{author}{\bibfnamefont{P.~A.} \bibnamefont{Parrilo}},
  \bibnamefont{and} \bibinfo{author}{\bibfnamefont{F.~M.}
  \bibnamefont{Spedalieri}}, \bibinfo{journal}{Phys. Rev. Lett.}
  \textbf{\bibinfo{volume}{88}}, \bibinfo{pages}{187904}
  (\bibinfo{year}{2002}), \eprint{quant-ph/0112007}.

\bibitem[{\citenamefont{Doherty et~al.}(2004)\citenamefont{Doherty, Parrilo,
  and Spedalieri}}]{DPS}
\bibinfo{author}{\bibfnamefont{A.~C.} \bibnamefont{Doherty}},
  \bibinfo{author}{\bibfnamefont{P.~A.} \bibnamefont{Parrilo}},
  \bibnamefont{and} \bibinfo{author}{\bibfnamefont{F.~M.}
  \bibnamefont{Spedalieri}}, \bibinfo{journal}{Phys. Rev. A}
  \textbf{\bibinfo{volume}{69}}, \bibinfo{pages}{022308}
  (\bibinfo{year}{2004}), \eprint{quant-ph/0308032}.

\bibitem[{\citenamefont{Eisert et~al.}(2004)\citenamefont{Eisert, Hyllus,
  G\"uhne, and Curty}}]{EHOC04}
\bibinfo{author}{\bibfnamefont{J.}~\bibnamefont{Eisert}},
  \bibinfo{author}{\bibfnamefont{P.}~\bibnamefont{Hyllus}},
  \bibinfo{author}{\bibfnamefont{O.}~\bibnamefont{G\"uhne}}, \bibnamefont{and}
  \bibinfo{author}{\bibfnamefont{M.}~\bibnamefont{Curty}},
  \bibinfo{journal}{Phys. Rev. A} \textbf{\bibinfo{volume}{70}},
  \bibinfo{pages}{062317} (\bibinfo{year}{2004}), \eprint{quant-ph/0407135}.

\bibitem[{\citenamefont{Mazziotti}(2006)}]{Maz06}
\bibinfo{author}{\bibfnamefont{D.~A.} \bibnamefont{Mazziotti}},
  \bibinfo{journal}{Acc. Chem. Res.} \textbf{\bibinfo{volume}{39}},
  \bibinfo{pages}{207} (\bibinfo{year}{2006}).

\bibitem[{\citenamefont{Hall}(2005)}]{Hall}
\bibinfo{author}{\bibfnamefont{W.}~\bibnamefont{Hall}}, \bibinfo{journal}{Phys.
  Rev. A} \textbf{\bibinfo{volume}{72}}, \bibinfo{pages}{022311}
  (\bibinfo{year}{2005}), \eprint{quant-ph/0504154}.

\bibitem[{\citenamefont{Han et~al.}(2005{\natexlab{a}})\citenamefont{Han,
  Zhang, and Guo}}]{HZG05}
\bibinfo{author}{\bibfnamefont{Y.-J.} \bibnamefont{Han}},
  \bibinfo{author}{\bibfnamefont{Y.-S.} \bibnamefont{Zhang}}, \bibnamefont{and}
  \bibinfo{author}{\bibfnamefont{G.-C.} \bibnamefont{Guo}},
  \bibinfo{journal}{Phys. Rev. A} \textbf{\bibinfo{volume}{72}},
  \bibinfo{pages}{054302} (\bibinfo{year}{2005}{\natexlab{a}}),
  \eprint{quant-ph/0403151}.

\bibitem[{\citenamefont{Higuchi et~al.}(2003)\citenamefont{Higuchi, Sudbery,
  and Szulc}}]{HSS03}
\bibinfo{author}{\bibfnamefont{A.}~\bibnamefont{Higuchi}},
  \bibinfo{author}{\bibfnamefont{A.}~\bibnamefont{Sudbery}}, \bibnamefont{and}
  \bibinfo{author}{\bibfnamefont{J.}~\bibnamefont{Szulc}},
  \bibinfo{journal}{Phys. Rev. Lett.} \textbf{\bibinfo{volume}{90}},
  \bibinfo{pages}{107902} (\bibinfo{year}{2003}), \eprint{quant-ph/0209085}.

\bibitem[{\citenamefont{Higuchi}(2003)}]{Hig03}
\bibinfo{author}{\bibfnamefont{A.}~\bibnamefont{Higuchi}}, \bibinfo{journal}{J.
  Math. Phys}  (\bibinfo{year}{2003}), \bibinfo{note}{to appear.},
  \eprint{quant-ph/0309186}.

\bibitem[{\citenamefont{Bravyi}(2004)}]{Brav04}
\bibinfo{author}{\bibfnamefont{S.}~\bibnamefont{Bravyi}},
  \bibinfo{journal}{Quantum Information and Computation}
  \textbf{\bibinfo{volume}{4}}, \bibinfo{pages}{12} (\bibinfo{year}{2004}),
  \eprint{quant-ph/0301014}.

\bibitem[{\citenamefont{Han et~al.}(2005{\natexlab{b}})\citenamefont{Han,
  Zhang, and Guo}}]{HZG04}
\bibinfo{author}{\bibfnamefont{Y.-J.} \bibnamefont{Han}},
  \bibinfo{author}{\bibfnamefont{Y.-S.} \bibnamefont{Zhang}}, \bibnamefont{and}
  \bibinfo{author}{\bibfnamefont{G.-C.} \bibnamefont{Guo}},
  \bibinfo{journal}{Phys. Rev. A} \textbf{\bibinfo{volume}{71}},
  \bibinfo{pages}{052306} (\bibinfo{year}{2005}{\natexlab{b}}),
  \eprint{quant-ph/0403151}.

\bibitem[{\citenamefont{Klaychko}(2004)}]{Kly04}
\bibinfo{author}{\bibfnamefont{A.}~\bibnamefont{Klaychko}}
  (\bibinfo{year}{2004}), \eprint{quant-ph/0409113}.

\bibitem[{\citenamefont{Daftuar and Hayden}(2005)}]{DH05}
\bibinfo{author}{\bibfnamefont{S.}~\bibnamefont{Daftuar}} \bibnamefont{and}
  \bibinfo{author}{\bibfnamefont{P.}~\bibnamefont{Hayden}},
  \bibinfo{journal}{Annals of Physics} \textbf{\bibinfo{volume}{315}},
  \bibinfo{pages}{80} (\bibinfo{year}{2005}), \eprint{quant-ph/0410052}.

\bibitem[{\citenamefont{Christandl and Mitchison}(2005)}]{CM05}
\bibinfo{author}{\bibfnamefont{M.}~\bibnamefont{Christandl}} \bibnamefont{and}
  \bibinfo{author}{\bibfnamefont{G.}~\bibnamefont{Mitchison}},
  \bibinfo{journal}{Comm. Math. Phys.} \textbf{\bibinfo{volume}{261}},
  \bibinfo{pages}{789} (\bibinfo{year}{2005}), \eprint{quant-ph/0409016}.

\bibitem[{\citenamefont{Jones and Linden}(2005)}]{JL05}
\bibinfo{author}{\bibfnamefont{N.~S.} \bibnamefont{Jones}} \bibnamefont{and}
  \bibinfo{author}{\bibfnamefont{N.}~\bibnamefont{Linden}},
  \bibinfo{journal}{Phys. Rev. A} \textbf{\bibinfo{volume}{71}},
  \bibinfo{pages}{012324} (\bibinfo{year}{2005}), \eprint{quant-ph/0407117}.

\bibitem[{\citenamefont{Coleman}(1963)}]{Col63}
\bibinfo{author}{\bibfnamefont{A.~J.} \bibnamefont{Coleman}},
  \bibinfo{journal}{Phys. Rev.} \textbf{\bibinfo{volume}{35}},
  \bibinfo{pages}{668} (\bibinfo{year}{1963}).

\bibitem[{\citenamefont{Liu et~al.}(2006)\citenamefont{Liu, Christandl, and
  Verstraete}}]{LCV06}
\bibinfo{author}{\bibfnamefont{Y.-K.} \bibnamefont{Liu}},
  \bibinfo{author}{\bibfnamefont{M.}~\bibnamefont{Christandl}},
  \bibnamefont{and}
  \bibinfo{author}{\bibfnamefont{F.}~\bibnamefont{Verstraete}}
  (\bibinfo{year}{2006}), \eprint{quant-ph/0609125}.

\bibitem[{\citenamefont{Sutherland}(1975)}]{SutBk}
\bibinfo{author}{\bibfnamefont{W.~A.} \bibnamefont{Sutherland}},
  \emph{\bibinfo{title}{Introduction to Metric and Topological Spaces}}
  (\bibinfo{publisher}{Oxford University Press}, \bibinfo{year}{1975}).

\bibitem[{\citenamefont{Edwards}(1965)}]{EdBk}
\bibinfo{author}{\bibfnamefont{R.~E.} \bibnamefont{Edwards}},
  \emph{\bibinfo{title}{Functional Analysis: Theory and Applications}}
  (\bibinfo{publisher}{Holt, Rinehart and Winston}, \bibinfo{year}{1965}).

\bibitem[{\citenamefont{Horodecki et~al.}(1996)\citenamefont{Horodecki,
  Horodecki, and Horodecki}}]{HHH96}
\bibinfo{author}{\bibfnamefont{M.}~\bibnamefont{Horodecki}},
  \bibinfo{author}{\bibfnamefont{P.}~\bibnamefont{Horodecki}},
  \bibnamefont{and}
  \bibinfo{author}{\bibfnamefont{R.}~\bibnamefont{Horodecki}},
  \bibinfo{journal}{Phys. Lett. A} \textbf{\bibinfo{volume}{223}},
  \bibinfo{pages}{1} (\bibinfo{year}{1996}), \eprint{quant-ph/9605038}.

\bibitem[{\citenamefont{Lewenstein et~al.}(2000)\citenamefont{Lewenstein,
  Kraus, Cirac, and Horodecki}}]{LKCH00}
\bibinfo{author}{\bibfnamefont{M.}~\bibnamefont{Lewenstein}},
  \bibinfo{author}{\bibfnamefont{B.}~\bibnamefont{Kraus}},
  \bibinfo{author}{\bibfnamefont{J.~I.} \bibnamefont{Cirac}}, \bibnamefont{and}
  \bibinfo{author}{\bibfnamefont{P.}~\bibnamefont{Horodecki}},
  \bibinfo{journal}{Phys. Rev. A} \textbf{\bibinfo{volume}{62}},
  \bibinfo{pages}{052310} (\bibinfo{year}{2000}), \eprint{quant-ph/0005014}.

\bibitem[{\citenamefont{Liu}(2006)}]{Liu06a}
\bibinfo{author}{\bibfnamefont{Y.-K.} \bibnamefont{Liu}}, in
  \emph{\bibinfo{booktitle}{Approximation, Randomization, and Combinatorial
  Optimization. Algorithms and Techniques}}, \bibinfo{organization}{Lecture
  Notes In Computer Science 4110} (\bibinfo{publisher}{Springer-Verlag
  (Germany)}, \bibinfo{year}{2006}), pp. \bibinfo{pages}{438--449},
  \eprint{quant-ph/0604166}.

\bibitem[{\citenamefont{Sturm}()}]{SEDUMI}
\bibinfo{author}{\bibfnamefont{J.}~\bibnamefont{Sturm}},
  \emph{\bibinfo{title}{Sedumi version 1.05}}, \bibinfo{howpublished}{Available
  at \texttt{http://sedumi.mcmaster.ca}}.

\bibitem[{\citenamefont{Aharonov and Naveh.}(2002)}]{AhN02}
\bibinfo{author}{\bibfnamefont{D.}~\bibnamefont{Aharonov}} \bibnamefont{and}
  \bibinfo{author}{\bibfnamefont{T.}~\bibnamefont{Naveh.}}
  (\bibinfo{year}{2002}), \eprint{quant-ph/0210077}.

\end{thebibliography}

\end{document}